\documentclass[runningheads]{llncs}

\usepackage[T1]{fontenc}
\def\doi#1{\href{https://doi.org/\detokenize{#1}}{\url{https://doi.org/\detokenize{#1}}}}
\usepackage{multirow}
\usepackage{tabularx}
\usepackage{amssymb}
\usepackage{amsmath}
\usepackage{graphicx}
\usepackage{amsfonts}
\usepackage{booktabs}
\usepackage{array}
\usepackage{siunitx}
\usepackage{cite}
\usepackage{hyperref}
\usepackage[misc]{ifsym}
\newcolumntype{C}[1]{>{\centering\let\newline\\\arraybackslash\hspace{0pt}}m{#1}}
\usepackage{listings}
\begin{document}
\title{Subject-Specific Lesion Generation and Pseudo-Healthy Synthesis for Multiple Sclerosis Brain Images}
\titlerunning{Subject-Specific Lesion Generation and Pseudo-Healthy Synthesis}
% If the paper title is too long for the running head, you can set
% an abbreviated paper title here
%
\author{Berke Doga Basaran\inst{1,2}\textsuperscript{(\Letter)} \orcidID{0000-0001-7235-3962} \and \\
Mengyun Qiao\inst{2} \orcidID{0000-0002-5157-1079} \and 
Paul M. Matthews\inst{3,4}\orcidID{0000-0002-1619-8328} \and
Wenjia Bai\inst{1,2,3}\orcidID{0000-0003-2943-7698}}
%
%index{Last Name, First Name}
%index{Last Name, First Name}
%index{Last Name, First Name}
%index{Last Name, First Name}

\authorrunning{B. D. Basaran et al.}
% First names are abbreviated in the running head.
% If there are more than two authors, 'et al.' is used.
%
\institute{Department of Computing, Imperial College London, London, UK \and
Data Science Institute, Imperial College London, London, UK \\
\email{bdb19@imperial.ac.uk} \and
Department of Brain Sciences, Imperial College London, London, UK \and
UK Dementia Research Institute, Imperial College London, London, UK \\
}
\maketitle           
\begin{abstract}
Understanding the intensity characteristics of brain lesions is key for defining image-based biomarkers in neurological studies and for predicting disease burden and outcome. In this work, we present a novel foreground-based generative method for modelling the local lesion characteristics that can both generate synthetic lesions on healthy images and synthesize subject-specific pseudo-healthy images from pathological images. Furthermore, the proposed method can be used as a data augmentation module to generate synthetic images for training brain image segmentation networks. Experiments on multiple sclerosis (MS) brain images acquired on magnetic resonance imaging (MRI) demonstrate that the proposed method can generate highly realistic pseudo-healthy and pseudo-pathological brain images. Data augmentation using the synthetic images improves the brain image segmentation performance compared to traditional data augmentation methods as well as a recent lesion-aware data augmentation technique, CarveMix. The code will be released at \href{https://github.com/dogabasaran/lesion-synthesis}{\texttt{https://github.com/dogabasaran/lesion-synthesis}}.

\keywords{Lesion generation \and Image synthesis \and Generative modelling \and Data augmentation \and Attention mechanisms.}
\end{abstract}

\section{Introduction}
Multiple sclerosis (MS) is a demyelinating neurodegenerative disease of the nervous system \cite{Ghasemi2017}. On brain magnetic resonance (MR) images such as fluid-attenuated inversion recovery (FLAIR), MS is often manifested as hyperintense lesions distributed mostly in the white matter \cite{Bakshi2001}. Identification and understanding the intensity patterns of these lesions thus plays a crucial role in diagnosing and tracking the progression of the disease. Being able to model the lesion intensity patterns also enables us to generate synthetic images for data augmentation purposes and to perform counter-factual inference to understand disease progression. Although many efforts have been dedicated in the past to the general image synthesis problem \cite{Huang2018, Karras2019, Reed2016}, little has been investigated for subject-specific MS lesion synthesis.

In this work, we propose a novel generative method to model the MS lesion characteristics. It is able to perform subject-specific lesion synthesis given a healthy image, as well as synthesise a pseudo-healthy image given a pathological one. An attention mechanism is designed so as to only modify the intensities in the lesion region while maintaining the structure in other brain regions. We demonstrate that the proposed method achieves high realism in both lesion generation and pseudo-healthy synthesis. In addition, using the synthetic images for training image segmentation networks, we can improve the lesion segmentation performance even in a low-data setting.

\subsection{Related Works}
\subsubsection{Brain Lesion Synthesis} Learning-based brain lesion synthesis has received a lot of attention recently \cite{Sun2020, Bowles2017}. Salem et al. uses an encoder-decoder U-Net structure to fuse lesion masks acquired from MS patients to healthy subjects \cite{Salem2019}. Bissoto et al., Jin et al., and Li et al. utilise generative adversarial networks (GANs) to synthesize new image samples \cite{Bissoto2018, Jin2021, Li2020a}. Yet,  \cite{Bissoto2018} and \cite{Li2020a} use semantic maps from pathological subjects to generate images, which lead to augmented images which have identical foreground shapes and labels to the samples that they are drawn from. \cite{Jin2021} uses a user-specified tumour mask to generate synthetic tumours on healthy scans. These methods do not generate diverse data samples with high foreground variance (lesion/tumour masks) which differ from already present samples in the training dataset. Reinhold et al. employs a casual model using variational autoencoders to generate lesions with predefined lesion load \cite{Reinhold2021}. As for subject-specific pathological to healthy image synthesis, known as pseudo-healthy synthesis, Xia et al. uses a framework that consists of a generator, a segmentor and a reconstructor, along with a mask discriminator which distinguishes segmented masks from pathology masks \cite{Xia2020}.

\subsubsection{Data Augmentation} With the assistance of data augmentation methods, deep learning has defined the  state-of-the-art for lesion segmentation \cite{Zeng2020}. Data augmentation has improved model generalisability by increasing dataset variance, and has alleviated some of the issues caused due to scarcity of labelled data. However, traditional data augmentation methods, which include flipping, rotating, scaling, intensity alterations, and elastic deformations, do not significantly alter the original image, e.g. the appearance of lesions. Zhang et al. proposed CarveMix, derived from CutMix \cite{Yun2019}, which utilises a lesion-aware mix-based method to combine carved lesion patches with non-lesion images \cite{Zhang2021}. Both of these methods directly employ foreground masks from other subjects to compose an augmented image with already present lesion shapes and characteristics. Current data augmentation methods for brain lesions do not drastically change the local features of lesions, and a technique which can realistically generate new lesions from a learnt distribution will increase training sample diversity and segmentation performance.

\subsection{Contributions} We present a novel brain lesion synthesis method for multiple sclerosis images. Our contributions are four-fold: (1) Differing from other techniques, our method is able to preserve subject identity and generate a new lesion foreground class without directly sampling a lesion mask from another subject. (2) An attention-based foreground discriminator is introduced to generate realistic images focusing at the lesion region. (3) The method is able to generate both synthetic lesions as well as pseudo-healthy images, achieved via a cyclic structure. This enables generating a diverse set of brain images for data augmentation purposes. (4) Our method is stochastic, and is able to generate multiple augmented images with different lesion appearances from a single input image.

\section{Methods}
The goal is to generate pathological images with lesions, given healthy images as input, while preserving the subject identity, and vice versa, create pseudo-healthy images without the lesions in the pathological subjects. Inspired by CycleGAN \cite{Zhu2017} and AttentionGAN \cite{Tang2019}, we adopt a two-branch architecture to learn both the lesion generation and pseudo-healthy synthesis tasks. We employ the attention and content decoders from \cite{Tang2019}, which provide maps on \textit{where} and \textit{what} the important features of the foreground and background classes of the image are, respectively. However, we encourage the foreground decoder to learn lesion specific features. In particular, we introduce a foreground-based discriminator to focus on the generation of the lesion region. We assume that two unpaired datasets are available for learning: a healthy dataset, $X_H = \{x_H\}$, and a pathological dataset with lesions or white matter hyperintensities (WMHs), $X_P = \{x_P\}$. Lower case $x_{H}$ and $x_{P}$ denote single image samples from these datasets. For the pathological dataset, a corresponding foreground dataset, $X_F = \{x_F\}$, is constructed using the lesion masks. Each foreground image is a masked lesion intensity image, generated by multiplying the original pathological image from $X_{P}$ with its respective lesion mask. 

\begin{figure} \centering
\includegraphics[width=.99\textwidth]{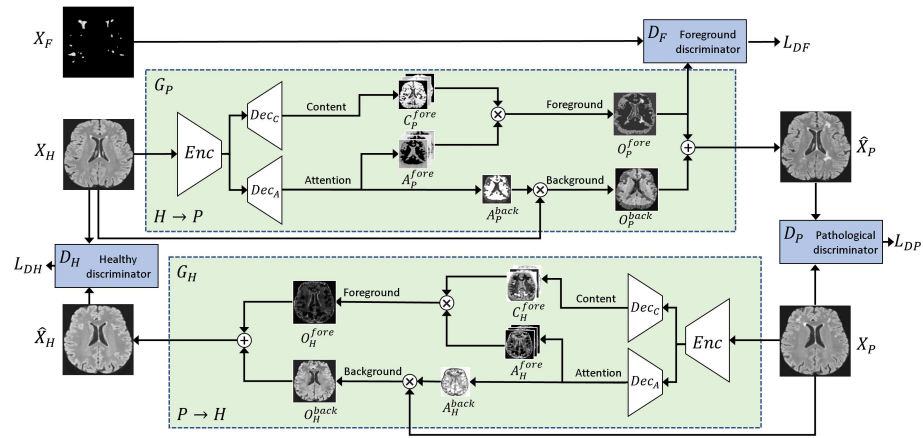}
\caption{Proposed framework for subject-specific lesion generation and pseudo-healthy synthesis of brain images, consisting of two generators, $G_{P}$ and $G_{H}$. See text for detail.} \label{framework}
\end{figure}

Figure \ref{framework} illustrates the proposed framework, which consists of two branches, one generator, $G_P$, for synthetic lesion generation and the other generator, $G_H$, for pseudo-healthy synthesis. We represent the lesion generation branch as $H \rightarrow P$, and the pseudo-healthy synthesis branch as $P \rightarrow H$, where $H$ denotes healthy images, and $P$ denotes pathological images. The lesion generator, $G_P$, takes in a healthy brain image, $x_H$, and generates a pathological image with lesions, $\hat{x}_P$, while also aiming to preserve subject identity. It is composed of an encoder, $Enc$, which maps the input image into a latent code $z$, a content decoder, $Dec_C$, that decodes $z$ into a foreground content map, $C^{fore}$, and an attention decoder, $Dec_A$, that decodes $z$ into the two attention maps, namely $A^{fore}$ for foreground and $A^{back}$ for background. Subsequently, the foreground content, $C^{fore}$, in our case the lesion-related regions, and attention map, $A^{fore}$, are fused to generate a foreground output, $O^{fore}$. The input image, $x_H$, and background attention, $A^{back}$, is fused to generate a background output, $O^{back}$. The summation of $O^{fore}$ and $O^{back}$ produces the generated synthetic image, $\hat{x}_P$. The pseudo-healthy synthesis branch, $G_H$, takes an image with lesions, $x_P$, as input and produces an pseudo-healthy image, $\hat{x}_H$. This branch is constructed in a similar way as the lesion generation branch.

\vspace{-.7cm}
\subsection{Generators}
The two generators, $G_P$ and $G_H$, are both constructed using the ResNet architecture with nine blocks. Each generator features an encoder with three downsampling paths, and two decoders, one for generating content masks, and one for attention masks. For a given input image, our model generates $n$ attention masks, 1 for background attention, $A^{back}$, and $n-1$ for foreground attention, $A^{fore}$, and also generates $n-1$ foreground content masks, $C^{fore}$. Here foreground broadly refers to regions relevant to lesions and background refers to the rest of the brain image. Producing multiple foreground attention masks allows lesions of varying shape and location characteristics to be decoded, thus producing synthetic images with different lesion distributions. Similar ideas have been explored in \cite{Chartsias2019}, where an input image is disentangled into multiple channels representing different anatomical structures. The synthetic image $G(x)$ is generated as,
\begin{equation}
\label{imagegenformula}
    G(x) =  C^{fore} * A^{fore} + x * A^{back}
\end{equation}
which combines the foreground content, $C^{fore}$, multiplied with their respective attention masks, $A^{fore}$, and the original input image, $x$, multiplied with the background attention mask, $A^{back}$. The separation of foreground and background allows us to explicitly focus on the lesions in the generating process by developing a lesion-aware foreground discriminator.

\subsection{Discriminators}
To generate realistic brain images, three discriminators are introduced. The pathological discriminator, $D_{P}$, distinguishes synthetic images with lesions from real pathological images. The healthy discriminator, $D_{H}$, distinguishes pseudo-healthy images from real healthy subjects. The third discriminator, $D_{F}$, which is lesion-aware, encourages the foreground (synthetic lesions) to look realistic compared to real lesion masks in the pathological dataset.
\vspace{-.1cm}

\subsection{Losses}
The total generator loss, $\mathcal{L}_{G_{total}}$, consists of three loss terms: generation loss, $\mathcal{L}_{G}$, cycle-consistency loss, $\mathcal{L}_{CC}$, and identity loss, $\mathcal{L}_{idt}$.  We employ the least squares GAN loss \cite{Mao2017} for the generation loss, also known as $\mathcal{L}_{2}$ loss. For pseudo-healthy synthesis ($P \rightarrow H$), the generation loss is formulated as, 
\begin{equation}
\label{ganloss1}
    \mathcal{L}_{G_{H}} = \mathbb{E}_{x\sim X_{P}} \left[ ( D_{H}(G_{H}(x))- 1)^{2}      \right]
\end{equation}
where $D_H$ denotes the healthy discriminator for the pseudo-healthy image $G_H(x)$. For lesion generation ($H \rightarrow P$), the generation loss is formulated as,
\begin{equation}
\label{ganloss2}
    \begin{split}
    \mathcal{L}_{G_{P}} = \mathbb{E}_{x\sim X_{H}} \left[ \left( \tfrac{1}{2}(D_{P}(G_{P}(x)) + D_{F}(O^{fore}_{P}))-1\right)^{2}\right]    
    \end{split}
\end{equation}
where $D_P$ denotes the pathological discriminator for the lesion image $G_P(x)$ and $D_F$ denotes the foreground discriminator for the foreground output $O^{fore}_{P}$, which penalises non-realistic lesion regions.

The cycle-consistency loss, $\mathcal{L}_{CC}$, evaluates a full cycle of an image through the network, $P \rightarrow H \rightarrow P$ or $H \rightarrow P \rightarrow H$, against its original input. We encourage the output of these cycles to be equivalent to the input as we wish to preserve subject identity. This also ensures that the generated synthetic lesions do not appear at unrealistic locations in the brain structure. The cycle-consistency loss, $\mathcal{L}_{CC}$, is formulated as,
\begin{equation}
\label{eq:cycleloss_hph}
\mathcal{L}_{CC} = \mathbb{E}_{x\sim X_{H}} \left[ \| G_{H}(G_{P}(x)) - x \|_{1} \lambda_{H} \right] + \mathbb{E}_{x\sim X_{P}} \left[ \| G_{P}(G_{H}(x)) - x \|_{1} \lambda_{P} \right]
\end{equation}

Finally, we incorporate the identity loss, $\mathcal{L}_{idt}$, with the objective of getting an identical image out of a generator if its healthiness is not changed, such that $G_{H}(x_{H})$ would return an image identical to $x_{H}$. We use the $\mathcal{L}_1$ loss for this task.
\begin{equation}
\label{idt_p}
\mathcal{L}_{idt} = \mathbb{E}_{x\sim X_{H}} \left[ \| G_{H}(x) - x \|_{1} \lambda_{H}\lambda_{idt} \right] + 
\mathbb{E}_{x\sim X_{P}} \left[ \| G_{P}(x) - x \|_{1}\lambda_{P}\lambda_{idt}\right] 
\end{equation}

We utilise hyperparameters $\lambda_{H}$, $\lambda_{P}$, and $\lambda_{idt}$ for weighting these losses. The total loss for the generator is summarized as,
\begin{equation}
\label{totalloss}
\mathcal{L}_{G_{total}} = \mathcal{L}_{G_{H}} + \mathcal{L}_{G_{P}} + \mathcal{L}_{CC} + \mathcal{L}_{idt}
\end{equation}

The discriminators aim to minimize the sum of the squared difference between predicted and expected values for real and synthetic images. We use the least squares loss for the three discriminators, $D_{H}$, $D_{P}$, $D_{F}$, formulated as, 
\begin{equation}
\label{hdiscrimloss}
\mathcal{L}_{D_{H}} = \mathbb{E}\left[ ( D_{H}(x_{H})- 1)^{2}  + (D_{H}(G_{H}(x_{P})))^{2}    \right]
\end{equation}
\begin{equation}
\label{pdiscrimloss}
\mathcal{L}_{D_{P}} = \mathbb{E}\left[ ( D_{P}(x_{P})- 1)^{2}  + (D_{P}(G_{P}(x_{H})))^{2}    \right]
\end{equation}
\begin{equation}
\label{fdiscrimloss}
\mathcal{L}_{D_{F}} = \mathbb{E}\left[ ( D_{F}(x_{F})- 1)^{2}  + (D_{F}(O^{fore}_{P}))^{2}    \right]
\end{equation}

\section{Experiments}
\subsection{Evaluation}
We evaluate the method performance in data augmentation using the synthetic data for downstream segmentation task as well as evaluate the image quality of synthetic data. To assess performance we conduct two experiments, employing our method as a data augmentation tool and comparing segmentation performance when trained on only synthetic data. To evaluate quality we ask three raters to assess healthiness and realism of generated images.

The proposed method is compared with traditional data augmentation (TDA) techniques which come default with nnU-Net \cite{Isensee2021}, including rotation, scaling, mirroring, elastic deformation, intensity perturbation, and simulation of low resolution. We also compare to a state-of-the-art lesion-aware augmentation method, CarveMix \cite{Zhang2021}. We utilise offline versions of our method and CarveMix to double the training dataset for those experiments. When performing data augmentation using the proposed method, lesion masks are required for the foreground discriminator. After lesion generation, we perform free-form deformation registration of the healthy subject image, $x_{H}$, to the corresponding synthesized pathological image, $\hat x_{P}$, threshold the difference of the two images and generate the lesion masks. We report the mean and variance of the Dice scores and Hausdorff distances of the segmentations.

\subsection{Implementation}
\subsubsection{GAN Network} We implement the generative model with a single channel input and single channel output. We set the generated mask number $n$ to 10, as our experiments show more variety of realistic lesion characteristics and intensities are synthesised with a higher $n$. Our generative framework method also features data augmentation in the form of sagittal mirroring and elastic deformation.

\subsubsection{Training} The proposed method is developed on PyTorch and trained on one NVIDIA GeForce RTX 3080. We use the Adam optimizer with an initial learning rate of 0.001 and 0.5 dropout. The network is trained for 400 epochs and with a linear decay of the learning rate after 200 epochs with a batch size of 1, due to GPU memory limit. We set the generator loss trade-off hyperparameters $\lambda_{H}$, $\lambda_{P}$, and $\lambda_{idt}$ to $10$, $10$, and $0.5$, respectively. For evaluating segmentation performance, nnU-Net is trained for 1,000 epochs using stochastic gradient descent.

\subsection{Data}
A private dataset is used for training the proposed method, which consists of FLAIR MRI scans of 120 subjects, including 60 healthy scans and 60 scans with MS lesions. For evaluation, we train the segmentation networks using the 2016 Multiple Sclerosis Lesion Segmentation dataset (MS2016) \cite{Commowick2018}, and test on the 2008 MICCAI MS Lesion Segmentation dataset (MS2008) \cite{Styner2008} and 2015 ISBI Longitudinal MS Lesion Segmentation dataset (ISBI2015) \cite{Carass2015}. We resample all images into $1\times1\times1$ mm$^3$ voxel spacing and extract the centre axial slice of each image. These images are reshaped into $256\times256$ dimensions to pass through the network. The private, MS2008, ISBI2015, and MS2016 datasets provide 120, 20, 21, and 15 2D FLAIR images, respectively.

\vspace{-.05cm}
\subsection{Results}
\subsubsection{Data Augmentation Performance}
We train twelve separate segmentation models using different dataset sizes and data augmentation methods. We use nnU-Net in 3D full resolution, utilizing 100\%, 53.3\%, 26.6\% and 12.3\% of the training dataset, corresponding to 15, 8, 4, and 2 subjects.

\begin{table*}[h!] \centering
\caption{Mean and standard deviations of the Dice scores ($\%$), at different sizes of training data. Best results are in bold. Asterisks indicate statistical significance ($^{*}$:~p$\leq$~0.05,  $^{**}$: p $\leq$ 0.001) when using a paired Student's \textit{t}-test compared to baselines.}
\label{table:segresults}
\begin{tabular*}{\textwidth}{l @{\extracolsep{\fill}} |cc|cc|cc} \toprule
      \multirow{2}{*}{Size} &  \multicolumn{2}{c|}{TDA} & \multicolumn{2}{c|}{TDA+CarveMix\cite{Zhang2021}} & \multicolumn{2}{c}{TDA+Proposed}   \\
      & MS2008 & ISBI2015 & MS2008 & ISBI2015 & MS2008 & ISBI2015 \\ 
     \midrule
       $100$\% & $33.06^{**}_{17.37}$ & $63.20_{16.70}$ & $32.32^{**}_{15.23}$ & $57.82_{16.56}$ & $\textbf{44.77}_{\textbf{15.87}}$ & $\textbf{63.90}_{\textbf{16.45}}$\\
       $53.3$\%  & $34.03_{17.83}$ & $58.84_{24.54}$ & $35.39_{17.36}$ & $61.25_{19.06}$ & $\textbf{39.70}_{\textbf{18.44}}$ & $\textbf{62.35}_{\textbf{13.21}}$\\
       $26.6$\%  & $31.55_{18.62}$ & $\textbf{64.43}_{\textbf{12.64}}$ & $\textbf{39.21}_{\textbf{18.86}}$ & $60.28_{21.22}$ & $34.58_{15.47}$ & $60.83_{10.03}$\\
       $13.3$\%  & $28.96_{23.55}$ &$ 37.34^{**}_{23.81}$ & $23.57{*}_{23.18}$ & $38.94^{**}_{21.81}$ & $\textbf{32.29}_{\textbf{16.16}}$ & $\textbf{56.64}_{\textbf{13.23}}$\\ \bottomrule
\end{tabular*}
\end{table*}

\begin{table*}[h!] \centering
\caption{Mean and standard deviations of the Hausdorff distances, at different sizes of training data. Best results are in bold. Asterisks indicate statistical significance ($^{*}$:~p$\leq$~0.05,  $^{**}$: p $\leq$ 0.01, $^{***}$: p $\leq$ 0.001) when using a paired Student's \textit{t}-test compared to baselines.}
\label{table:hd}
\begin{tabular*}{\textwidth}{l @{\extracolsep{\fill}} |cc|cc|cc} \toprule
      \multirow{2}{*}{Size} &  \multicolumn{2}{c|}{TDA} & \multicolumn{2}{c|}{TDA+CarveMix \cite{Zhang2021}} & \multicolumn{2}{c}{TDA+Proposed}   \\
      & MS2008 & ISBI2015 & MS2008 & ISBI2015 & MS2008 & ISBI2015 \\ 
     \midrule
       $100$\% & $3.52^{*}_{1.61}$ & $3.35^{*}_{0.94}$ & $3.24_{1.09}$ & $3.33_{0.89}$ & $\textbf{3.09}_{\textbf{{1.54}}}$ & $\textbf{3.02}_{\textbf{0.80}}$\\
       $53.3$\%  & $3.16^{**}_{1.10}$ & $3.12_{0.93}$ & $3.30^{***}_{1.26}$ & $3.28_{1.00}$ & $\textbf{2.62}_{\textbf{0.98}}$ & $\textbf{3.07}_{\textbf{1.08}}$\\
       $26.6$\%  & $3.12_{0.89}$ & $\textbf{3.38}_{\textbf{1.01}}$ & $\textbf{{3.08}}_{\textbf{1.01}}$ & $3.37_{0.94}$ & $3.09_{0.72}$ & $3.86_{1.61}$\\
       $13.3$\%  & $2.92_{1.13}$ & $3.86^{**}_{1.62}$ & $2.95_{1.03}$ & $3.81^{**}_{1.62}$ & $\textbf{2.90}_{\textbf{0.69}}$ & $\textbf{3.33}_{\textbf{1.17}}$\\ \bottomrule
\end{tabular*}
\end{table*}

\begin{figure}[h!]
\centering
\includegraphics[width=.77\textwidth]{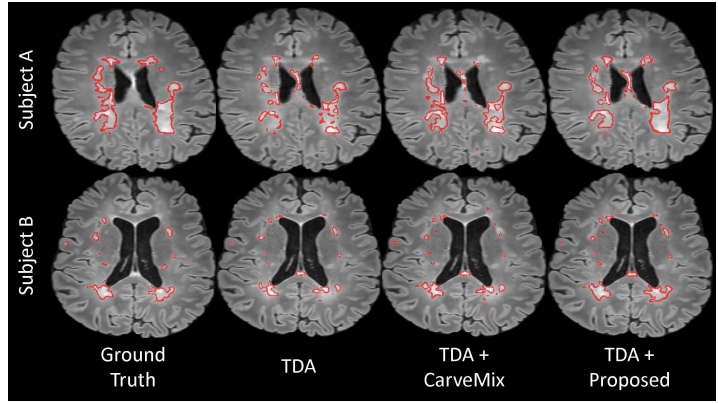}
\caption{Comparison of segmentation performance using different data augmentation techniques. Lesions are contoured in red. \textbf{Left to right:} Test subjects with ground truth lesion masks; segmentation using traditional data augmentation (TDA); segmentation using TDA+CarveMix \cite{Zhang2021}; segmentation using TDA+proposed.} \label{fig:seg}
\end{figure}

\begin{figure}[h!] \centering
\includegraphics[width=.8\textwidth]{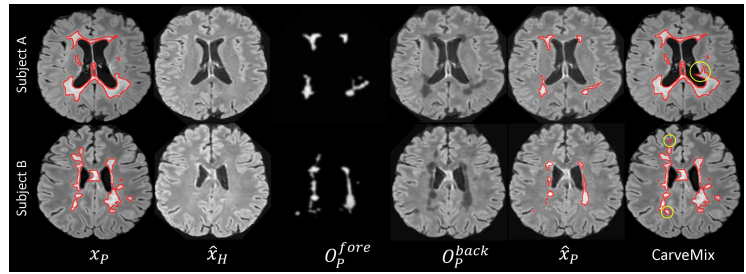}
\caption{Different stages of the proposed method. Lesions are contoured in red. \textbf{Left to right}: Real pathological image, $x_{P}$; pseudo-healthy image $\hat x_{H}$; foreground output for lesion synthesis on pseudo-healthy image $O^{fore}_{P}$; background output for lesion synthesis $O^{back}_{P}$; the synthetic pathological image generated from the pseudo-healthy image $\hat x_{P}$; augmented image using CarveMix, unrealistic lesions circled in yellow.} 
\label{fig:differentstages}
\end{figure}

Table \ref{table:segresults} indicates that the average Dice score is increased using the proposed method in three of the four dataset sizes when compared to TDA and CarveMix, while also decreasing the standard deviation in most cases. We further report the mean and variance of the Hausdorff distances in Table \ref{table:hd}. Figure \ref{fig:seg} visualises the segmentation performance of our method against the benchmark. Figure \ref{fig:differentstages} demonstrates different stages of the proposed method and augmentation examples from CarveMix. \cite{Zhang2021}. 

\begin{figure}[h!]
\centering
\includegraphics[width=.9\textwidth]{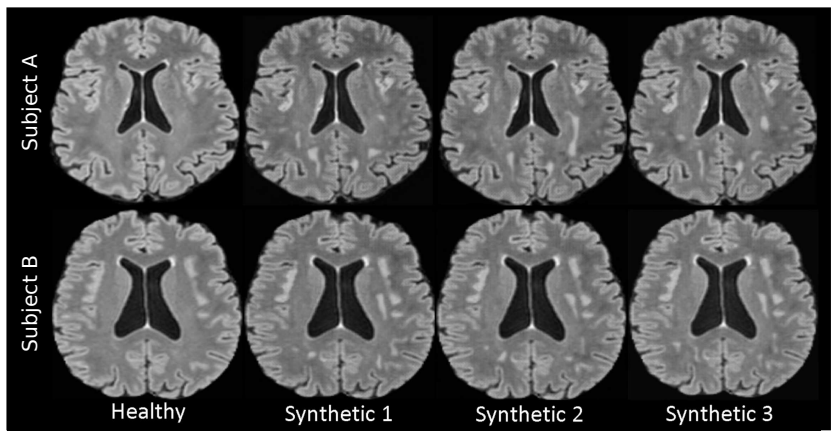}
\caption{Setting the generative network's dropout level to 0.5 at training and inference allows for generating a diverse set of synthetic images (second to fourth columns) given a single image as input (first column).} \label{fig:multifake}
\end{figure}

The generator of the method is stochastic, which employs dropout to generate multiple diverse augmented images for each given sample, shown in Figure \ref{fig:multifake}. This increases the diversity of training image samples.

\subsubsection{Synthetic data segmentation performance} 
We compare segmentation performance when training a model using a fully synthetic dataset against a model trained on a real dataset. We generate 15 synthetic images from the MS2016 dataset and test on the MS2008 and ISBI2015 datasets. We compare our results to nnU-Net with TDA at 100\% dataset size, as per in Table \ref{table:segresults}. Results in Table \ref{table:synthsegresults} suggest a model trained on our generated data performs comparable to a model trained on a real dataset.

\begin{table*}[h!] \centering
\caption{Segmentation performance using fully synthetic data for model training, compared to using real data. Mean and standard deviations of the Dice scores (\%) are reported with best results in bold.}
\label{table:synthsegresults}
\begin{tabular*}{.85\textwidth}{c @{\extracolsep{\fill}} cc} \toprule
     {Test data} &  {Real training data} & {Synthetic training data}   \\
     \midrule
       MS2008 & $33.06_{17.37}$ & $\textbf{35.74}_{\textbf{19.85}}$\\ 
       ISBI2015  & $\textbf{63.20}_{\textbf{16.70}}$ & $52.72_{18.61}$ \\ \bottomrule
\end{tabular*}
\end{table*}

\newpage
\subsubsection{Expert Evaluation}
In order to assess quality of generated images, we ask three human raters to evaluate images in healthiness and realism. We define "healthiness" as a brain MR image with no visible pathological features, and "realism" as an image which possesses the same quality and authentic look as one obtained by a scanner. We collate 50 real healthy, 50 real pathological, 50 pseudo-healthy, and 50 pseudo-pathological images. Raters are asked to classify each image as healthy (1) or pathological (0), and real (1) or fake (0). We produce the results for the real images as a benchmark to compare our generated images. Table \ref{table:raterresults} shows that the generated pseudo-healthy and pseudo-pathological images closely follow real images for the assessed metrics. 

\begin{table}[h!] \centering
\caption{Rater classification of real and generated images on realism and healthiness. Asterisk (*) indicates that a lower score is better.}
\label{table:raterresults}
    \begin{minipage}{.49\textwidth}
      \centering
      \begin{tabular}{p{3cm}C{1cm}c}
        \toprule \multicolumn{3}{c}{Realism}\\ 
        \midrule &  Real & Proposed \\
        \midrule Healthy images & 85\% & 82\%\\
        Pathological images & 84\% & 82\% \\
        \bottomrule
      \end{tabular}
    \end{minipage}
    \begin{minipage}{.49\textwidth}
      \centering
      \begin{tabular}{p{3cm}C{1cm}c}
        \toprule \multicolumn{3}{c}{Healthiness}\\ 
        \midrule & Real & Proposed\\
        \midrule Healthy images & 85\% & 83\%\\
        Pathological images*  & 5\% & 1\% \\
        \bottomrule
      \end{tabular}
    \end{minipage}
  \end{table}

\subsubsection{Correlation with Radiological Findings} 
\begin{figure}[h!]
    \centering
    \includegraphics[width=0.9\textwidth]{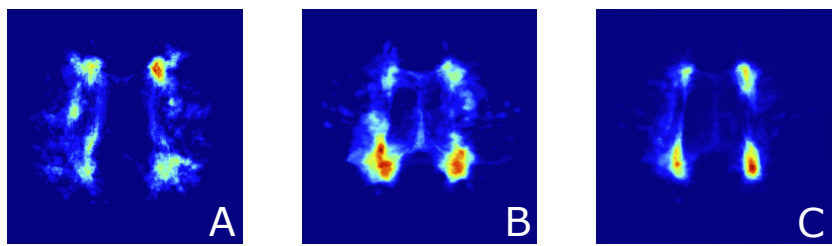}
    \caption{Comparison of spatial heatmaps of real lesions and synthetic lesions. \textbf{A:} Real lesions from private dataset (200 subjects). \textbf{B:} Real lesions in MS2016 dataset (60 subjects). \textbf{C}: Synthetic lesions generated using the proposed method by training on the MS2016 dataset (200 masks).}
    \label{fig:heatmap}
\end{figure}

The spatial heatmap of the synthetic lesions closely resemble those of the real lesions from clinical data, including the private dataset and the MS2016 dataset. A comparison between the spatial heatmap of synthetic lesions and real lesions is visualised in Figure \ref{fig:heatmap}.

\begin{figure}[h!]
\centering
\includegraphics[width=.75\textwidth]{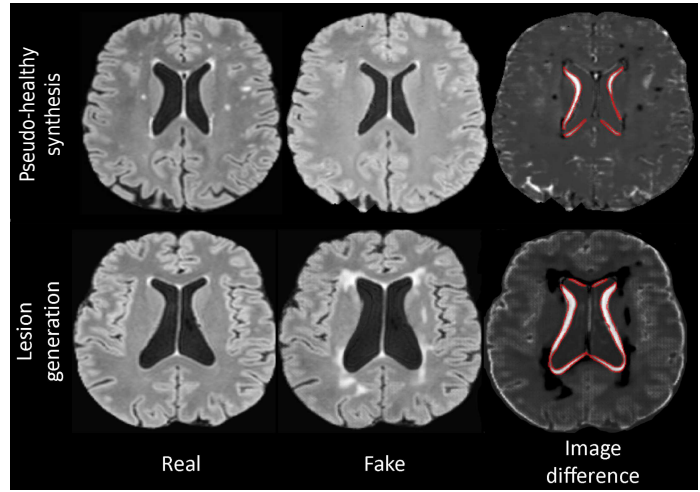}
\caption{Apart from generating lesions or pseudo-healthy synthesis, the proposed method also changes the ventricular volume during image generation, which is consistent with clinical literature. \textbf{Top:} Pseudo-healthy synthesis leads to decrease in ventricular volume, delineated in red in the difference image. \textbf{Bottom:} Lesion generation leads to an increase in ventricular volume, delineated in red in the difference image.} \label{fig:ventricles}
\end{figure}

Furthermore, generated pseudo-pathological images show an increase in the ventricular volume, while pseudo-healthy images tend to have a decreased ventricular volume. The difference in ventricular volume is shown in Figure \ref{fig:ventricles}. The change in the ventricular volume is consistent with literature \cite{Dalton693, Dalton2002, Guptha2002, Luxenberg1987}.

\section{Conclusion}
We have proposed a novel method for generating lesions and pseudo-healthy images while preserving subject-specific features, providing a useful tool for data augmentation in brain image analysis to complement real patient datasets. It is lesion-aware in the generating process. In quantitative evaluation, the method improves lesion segmentation performance in downstream tasks. In our human rater assessment, the generated images achieved a high realism score close to the real images, and generated lesions follow a spatial distribution consistent with real clinical datasets.

\subsubsection{Acknowledgements} This work is supported by the UKRI CDT in AI for Healthcare \href{http://ai4health.io}{http://ai4health.io} (Grant No. EP/S023283/1).

%
% ---- Bibliography ----
%
\newpage
\bibliographystyle{splncs04}
\bibliography{refs}
\end{document}